\documentclass[aps,preprint,showpacs, showkeys]{revtex4}%
\usepackage{amsfonts}
\usepackage{amsmath}
\usepackage{amssymb}
\usepackage{graphicx}%
\providecommand{\U}[1]{\protect\rule{.1in}{.1in}}
\begin{document}
\title[ ]{Will the $\mathcal{PT}$-symmetric and Non-Hermitian $\phi^{4}$ Theory Solve
the Hierarchy and Triviality Problems in the Standard Model?}
\author{Abouzeid M. Shalaby\footnote{E-mail:amshalab@ mans.eg.edu}}
\affiliation{Physics Department, Faculty of Science, Mansoura University, Egypt.}

\begin{abstract}
Very recently, the Large Hadron Collider was turned on. There, the experiments
are aiming to test different scenarios for elementary particles interactions
from SUSY, Extra dimensions to others. In fact, SUSY was invented to kill the
conceptual problems existing in the conventional Standard model \textit{i.e.}
the Hierarchy problem. However, in recent years, certain theories which was
rejected in the past like the wrong sign $\phi^{4}$ theory is now well known
to be a truly physically acceptable theory. Here, we analyze the
renormalization group flow of the different parameters in the theory. We find
that, rather than the conventional $\phi^{4}$ theory and because the theory is
asymptotically free, it does not sufferer from the catastrophic blow up of the
mass parameter for UV scales. This feature greatly recommend that this theory
is a plausible candidate to play the role of the $SU(2)\times U(1)$ symmetry
breaking in the standard model. If this picture will agree with the
experimental results from LHC, then the Higgs particle was massless in the far
past. Moreover, the cosmological constant problem as a Hierarchy problem may
be solved too. However, when trying to calculate the metric operator, we
expected that the equivalent Hermitian theory may attain non-remormalizabe
terms and thus the theory is incalculable. In fact, we show that there exists
a Hermitian and renormalizable equivalent theory though of indefinite metric.
We reformulated the Klein-Gordon equation in such a way that the wrong sign in
the propagator attains a new physical meaning that the Higgs particle is made
from exotic matter.

\end{abstract}
\maketitle

The origin of mass of the building blocks of our universe impresses the
nowadays physicists \cite{mass}. The reason behind that is the existence of
conceptual problems in the Standard model. Such conceptual problems occur
because of our ultimate hope to unify all the existent forces. The unification
of the electromagnetic, weak and strong interactions introduces the energy
scale $10^{16}$ $GeV$, the scale at which the three couplings be the same (in
the supersymmetry (SUSY) regime). However, we have another two interesting
energy scales, the electroweak scale and the Planck scale. In fact, the scalar
Higgs does possess two problems regarding the flow of the parameters either
for small or high energy scales. For low energy scales, the theory is trivial
while if one tried to let the dimensionful parameters to flow to a very high
energy scales, one gets big values for those parameters which will violate the
observation. The big values for the dimensionful parameters are well known as
gauge Hierarchy problem.

There exists more than one algorithm suggested to solve the Hierarchy problem.
For instance, SUSY has been introduced with the aim to have natural
cancellation in the dimensionful parameters that turned those parameters
protected against perturbations even for very high energy scales \cite{susy}.
However, it introduces an upper limit to the Higgs mass by 130 $GeV$ and some
of its mass spectra are of one $TeV$ which make this theory under the fire of
the LHC experiments test by 2008. Another algorithm for the solution of the
Hierarchy problem is to consider the Higgs particle as a composite state bound
by a new set of interactions (Technicolors)\cite{tecni}. However, the
technicolor model is strongly constrained from precision tests of electroweak
theory at LEP and SLC experiments \cite{tecni1} . Also, this algorithm has
mass spectra of about one $TeV$ and it is under the direct test of the LHC
experiments. Once more, a recent algorithm is suggested for which the
$SU(2)\times U(1)$ symmetry is broken via the compactification of an extra
dimension. In fact, particles in this model attain masses through the
expectation value of the fifth ( for instance) component of the gauge field.
However, to some physicists, the digestion of extra dimension is not that easy
and can be accepted to them at most as a mathematical modeling to the problem.
In this letter, we analyze a modified algorithm which we think it can solve
the famous triviality as well as the Hierarchy problems. In this algorithm, we
revisit a previously rejected theories because they are non-Hermitian but in
view of the recent discoveries of the reality of some class of non-Hermitian
models ($\mathcal{PT}$-symmetric), there exists no reason to prevent them from
playing in the scene. We think that this is fair and the final word about the
correct scenario may come from the analysis of the experimental data from the LHC.

In 1998, Carl Bender and Stefan Boettcher have shown that a class of
Non-Hermitian but $\mathcal{PT}$-symmetric Hamiltonians have real spectra
\cite{bendr}. This discovery led us to reinvestigate the non-Hermitian quantum
field models like the $(-g\phi^{4})$ and $\left(  g\phi^{4}+h\phi^{6}\right)
$ scalar theories \cite{abophi41p1,abophi61p1}. In the first model, we
realized that, rather than the corresponding Hermitian model $(g\phi^{4})$,
the vacuum energy is tiny for a wide range of energy scales. Besides, the
vacuum energy is real and in fact, it is easy to show that the $PT$ symmetry
is exact which verify that not only the lowest energy but the full energy
spectrum is real \cite{aboinner}. However, what makes this field theory very
impressive is that it is asymptotically free \cite{Symanzik,bendf,Frieder}.
Moreover, we discovered that the Hermitian model $\left(  g\phi^{4}+h\phi
^{6}\right)  $ can have a transition to a phase for which the theory is
physically acceptable though non-Hermitian \cite{abophi61p1}. In fact, we
asserted that this model has a very interesting property namely,
matter-antimatter asymmetry, which is the first time to be obtained in a
spontaneously symmetry breaking regime. Relying on such interesting properties
of $\mathcal{PT}$-symmetric and non-Hermitian models, one may ask if such new
subject can help in solving the above mentioned problems in the standard
model. In this letter, we answer this question by renormalization group
analysis of the $(-\frac{\lambda}{4!}\phi^{4})$ scalar field model in $3+1$
dimension. In fact, the idea we relied on is that the theory is asymptotically
free and conclusions drawn from this model can be generalized to the more
reliable complex scalar field (doublets) that is used to break the
$SU(2)\times U(1)$ symmetry in the standard model.

\begin{figure}[ptbh]
\begin{center}
\includegraphics{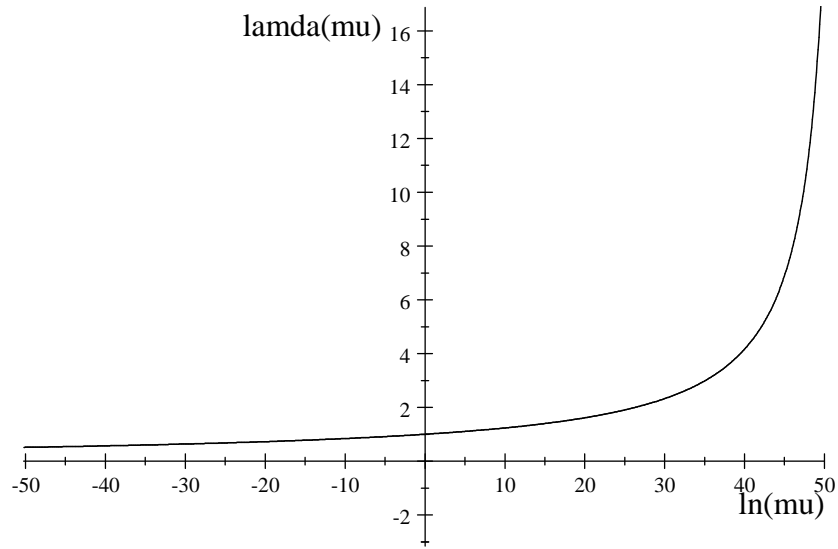}
\end{center}
\caption{The dependence of the renormalized coupling $\lambda(\mu)$ on the
mass scale $\mu$ for the Hermitian $\phi^{4}$ theory. In this figure, the
theory is shown to be trivial for $IR$ scales, while the coupling blows up for
$UV$ scales which means that the mass parameter will explode
non-logarithmically at $UV$ scales causing the Hierarchy problem.}%
\label{lamdah}%
\end{figure}

\begin{figure}[ptbh]
\begin{center}
\includegraphics{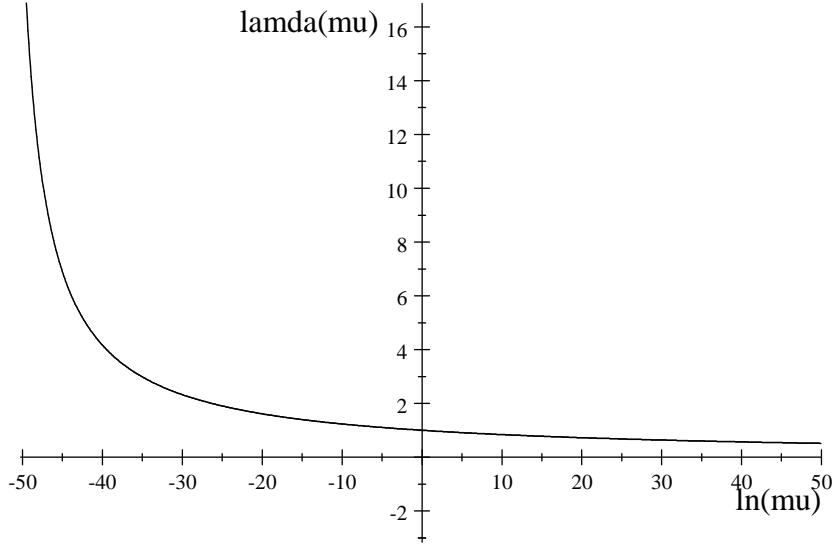}
\end{center}
\caption{The dependence of the renormalized coupling $\lambda(\mu)$on the mass
scale $\mu$ for the non-Hermitian $\phi^{4}$ theory. One can realize that the
theory is non-trivial as well as asymptotically free. Accordingly, the mass is
finite for ultra $UV$ scales.}%
\label{lamdanonh}%
\end{figure}

In the $\phi^{4}$ model in $3+1$ dimension, which is used to break the
$SU(2)\times U(1)$ symmetry in the standard model, up to one loop, we realize
that the mass term receives a correction of the form;%

\begin{equation}
M_{H}^{2}=M_{0}^{2}+\frac{3\lambda\Lambda^{2}}{8\pi^{2}},
\end{equation}
where the $M_{H}^{2}$ and $M_{0}^{2}$ are the renormalized and the bare mass
squared of the Higgs particle while $\lambda$ is the coupling constant. In
fact, the appearance of the momentum cutoff $\Lambda$ is the reason behind the
Hierarchy problem, which leads to the introduction of the supersymmetry where
the cutoff $\Lambda$ from a Boson and a Fermion loops cancels. Instated of the
SUSY additive cancellation, one can guess a multiplication softening of the
Hierarchy problem. By this we mean, if we have an asymptotically free Higgs
particle, when $\Lambda$ is very high, the coupling is very small and thus one
may expect that the perturbative correction $\frac{3\lambda\Lambda^{2}}%
{8\pi^{2}}$ stays small. To test this idea, consider the renormalization group
functions of the $\mathcal{PT}$-symmetric and non-Hermitian $(-\frac{\lambda
}{4!}\phi^{4})$ scalar field model in $3+1$ dimensions up to second order in
the coupling;%

\begin{align}
\beta\left(  \lambda\right)   &  =-\frac{3\lambda^{2}}{\left(  4\pi\right)
^{2}},\\
\gamma_{m}\left(  \lambda\right)   &  =\frac{-\lambda}{\left(  4\pi\right)
^{2}}-\frac{5}{6}\left(  \frac{\lambda}{\left(  4\pi\right)  ^{2}}\right)
^{2},
\end{align}
where $\beta\left(  \lambda\right)  =\mu\frac{d\lambda}{d\mu}$ and $\gamma
_{m}\left(  \lambda\right)  =\frac{\mu}{m}\frac{dm}{d\mu}$ are the well known
renormalization group functions for the flow of the coupling and the mass
parameters. Accordingly, the mass parameter can be given by;%
\begin{align}
m^{2}\left(  \mu\right)   &  =m^{2}\left(  \mu_{0}\right)  \exp\left(
\int_{\lambda_{0}}^{\lambda_{\mu}}\frac{\gamma_{m}\left(  \lambda\right)
}{\beta\left(  \lambda\right)  }d\lambda\right) \\
&  =m^{2}\left(  \mu_{0}\right)  \exp\left(  -\frac{1}{288}\frac
{-5\lambda_{\mu}-96\left(  \ln\frac{\lambda_{\mu}}{\lambda_{0}}\right)
\pi^{2}+5\lambda_{0}}{\pi^{2}}\allowbreak\right)  .
\end{align}
In fact, because $\beta\left(  \lambda\right)  $ is negative it will drag the
coupling to the origin as $\mu$ goes to higher values. This behavior is well
known as the asymptotic freedom. Accordingly, the Higgs mass will go also to
very small values at high energy scales. Thus, if this picture is the
successful one in view of the coming analysis from LHC, not only was the
quarks, leptons, vector Bosons were massless in the far past but also the
Higgs particle was massless. To make the difference between the features of
the Hermitian $\phi^{4}$ and the $\mathcal{PT}$-symmetric $\phi^{4}$ more
illustrative, we plotted the behavior of the coupling constant as a function
of the logarithm of the mass scale $\mu$ in Fig. \ref{lamdah} and
Fig.\ref{lamdanonh}, respectively. One can easily realize from the figures
that the $\mathcal{PT}$-symmetric $\phi^{4}$ theory is non-trivial and
asymptotically free while the Hermitian $\phi^{4}$ is both trivial and the
coupling blows up for $UV$ scales which causes the hierarchy problem. In fact,
the main message we want to spreed in this letter is that (i) there is no
gauge Hierarchy problem with the non-Hermitian and $\mathcal{PT}$-symmetric
Higgs mechanism, provided that the contribution to the renormalization group
functions from other sectors in the standard model will not affect the
asymptotic freedom property of the scalar sector \ (ii) the model is
non-trivial (iii) the technical problem concerning the remedy of the
indefinite metric of the theory in the conventional Hilbert space may be
solved via a simple Bogoliubov transformation for which the new representation
is Hermitian and thus the theory is unitary though of indefinite norm which we
will try to attribute it to the existence of a new physical meaning. In fact,
it seems that we are obligated to follow that route as the existing regimes
for handling the $\mathcal{PT}$-symmetric theories introduces a metric
operator of the form $\eta_{+}=\int d^{3}x e^{-Q(x)}$ which is expected to
introduce non-renormalizable terms in the equivalent Hermitian theory, in case
we are able to calculate the metric operator for this theory in $3+1$
dimensions \footnote{Up to the best our knowledge, the metric operator for the
$\mathcal{PT}$-symmetric ($-\phi^{4}$) theory has never been obtained
before.}. We will investigate this point later on in this work.

Now, we may speculate about if this is the correct picture, why we did not
discovered the Higgs particle yet? The answer to this question may be that, in
this picture the Higgs particle is a strongly interacting particle and one can
not isolate a single Higgs, the same way of behavior of quarks and gluons.
Then, one may instead talk about bound states of Higgs which we call it Higgs
balls ( like glue balls). Thus, according to this picture, the discovery of
the Higgs is not a matter of building bigger and bigger machines for the sake
of higher and higher energies but a matter of our understanding of the nature
of the Higgs particle.

Although of the above mentioned amazing features of the non-Hermitian and
$\mathcal{PT}$-symmetric $\phi^{4}$ theory toward the solution of the genuine
problems in the standard model, there exists a technical problem concerning
the expected negative norm of the theory. In fact, in the Hilbert space with
the Dirac sense inner product operation, the theory have a positive norm but
unitarity is violated. This led Bender \textit{et.al} to suggest the $PT$
inner product of the form \cite{bend2005};%

\[
\langle A|B\rangle_{PT}=(PT|A\rangle)^{T}|B\rangle.
\]
This suggestion preserves unitarity but the Hilbert space with the $PT$ inner
product has an indefinite metric. Again, this led Bender et.al. to introduce
what is called the $C$ operator and the Hilbert space with the $CPT$ inner
product preserves unitarity as well as having a positive definite metric.
However, the calculation of the $C$ operator for the model under consideration
is not that easy and will turn the computation cumbersome. Although this is a
technical problem and not a conceptual one, up to the best of our knowledge,
no body has obtained the $C$ operator for the non-Hermitian and $\mathcal{PT}%
$-symmetric $\phi^{4}$ theory. However, another (equivalent) regime to cure
the indefinite metric problem has been suggested by Mostafazadeh \cite{zadah}.
What is important to us here from the Mostafazadeh regime is that the
non-Hermitian form of the Hamiltonian is nothing but another representation of
an equivalent Hermitian representation and one can (in principle) switch
between the two representations via a similarity transformation. However, the
metric operator in the Mostafazadeh regime is hard to get in the non-real line
theories especially in quantum field models. In fact, both Bender and
Mostafazadeh regimes will lead to a dynamical Feynmann rules in the sense that
the Feynmann rules will change from order to order because of the new
operators introduced to the theory in the definition of the positive definite
inner product. Accordingly, one may ask if there exist a simple similarity
transformation by which one can obtain an equivalent Hermitian theory and thus
having a Hilbert space with the conventional Dirac sense inner product. We
will try to answer this question in the following part of the work.

The reality of the spectrum of a non-Hermitian $\mathcal{PT}$-symmetric
Hamiltonian demands an exact $PT$ symmetry in the sense that all the wave
functions respect the $PT$ symmetry in the same footing as the Hamiltonian
itself. In the above discussions, we did not check if the states have an exact
$PT$ symmetry or not. To show that, let $|n(k)\rangle$ is a state consisting
of $n(k)$ identical particles with momentum $k$. Because the field is assumed
to transform as a pseudo scalar under $PT$ transformations we get the
transformation of the creation and annihilation operators as;
\begin{equation}
PTa(PT)^{-1}=PT\left(  i\int dx\left(  \exp\left(  -ikx\right)  \pi
-\phi\overleftrightarrow{\left(  \frac{\partial}{\partial t}\right)  }%
\exp\left(  ikx\right)  \right)  \right)  (PT)^{-1}=-a,
\end{equation}
where $\overleftrightarrow{\left(  \frac{\partial}{\partial t}\right)
}\left(  AB\right)  =A\frac{\partial B}{\partial t}-\frac{\partial A}{\partial
t}B$.\newline Also, $PTa^{\dagger}(PT)^{-1}=-a^{\dagger}$.\newline Since,
\begin{align}
|n(k)\rangle &  =\frac{a^{\dagger n(k)}(k)|0\rangle}{\sqrt{n(k)}!},\\
PT|n(k)\rangle &  =\left(  -1\right)  ^{n(k)}|n(k)\rangle,
\end{align}
where we observe that $\mathcal{PT}$-symmetry is broken. To keep the $PT$
symmetry unbroken, we add the famous extra $i^{n}$ factor to the states in the
following way;
\begin{align}
|n(k)\rangle &  =\frac{i^{n(k)}a^{\dagger n(k)}(k)|0\rangle}{\sqrt{n(k)}!},\\
PT|n(k)\rangle &  =|n(k)\rangle.
\end{align}
Also, a state consisting of many particles of different momenta, can be
represented by
\begin{align}
|n(k_{1})n(k_{1})n(k_{3})n(k_{4})....n(k_{m})\rangle &  =\prod_{j=1}%
^{j=m}\frac{i^{n(k_{j})}a^{\dagger n(k_{j})}(k_{j})|0\rangle}{\sqrt{n(k_{j}%
)}!},\\
PT|n(k)\rangle &  =|n(k)\rangle.
\end{align}
In this way one can build up states which are $\mathcal{PT}$-symmetric too and
thus the $\mathcal{PT}$ symmetry is exact which is an essential requirement
for the reality of the $\mathcal{PT}$-symmetric Hamiltonian.

Now consider the Hamiltonian model of the form;%
\[
H=H=\frac{1}{2}\left(  \left(  \nabla\phi\right)  ^{2}+\pi^{2}+m^{2}\phi
^{2}\right)  -\frac{\lambda}{4!}\phi^{4},
\]
where $\pi$ is the conjugate momentum. Making the field shift $\phi=\psi+B$;
one can rewrite the Hamiltonian density in the form%

\begin{equation}
H=H_{0}+H_{1}+H_{I}+E, \label{quasi}%
\end{equation}
where
\begin{align*}
H_{0}  &  =\frac{1}{2}\left(  \left(  \nabla\psi\right)  ^{2}+\Pi^{2}%
+M^{2}\psi^{2}\right)  ,\\
H_{1}  &  =\left(  m^{2}-\frac{\lambda}{6}B^{2}\right)  B\psi,\\
H_{I}  &  =\frac{-\lambda}{4!}\left(  \psi^{4}+4B\psi^{3}\right)  ,\\
E  &  =\frac{1}{2}m^{2}B^{2}-\frac{\lambda}{4!}B^{4},
\end{align*}
where $B$ is the vacuum condensate, $\Pi=\overset{\cdot}{\psi}$ and
$M^{2}=m^{2}-\frac{\lambda}{8}B^{2}$. A well known condition for the effective
potential $E$ is to satisfy the equality $\frac{\partial E}{\partial B}=0$.
Accordingly, the term $H_{1}$ is ought to be zero. Thus, the quasi-particle
Hamiltonian in Eq.(\ref{quasi}) has the form
\begin{equation}
H=\int d^{3}x\left(  \frac{1}{2}\left(  \left(  \nabla\psi\right)  ^{2}%
+\Pi^{2}+M^{2}\psi^{2}\right)  -i\frac{\lambda\left\vert B\right\vert }{6}%
\psi^{3}-\frac{\lambda}{4!}\psi^{4}\right)  , \label{quasi2}%
\end{equation}
where we used the fact that the vacuum condensate of this theory is pure
imaginary \cite{abophi41p1,onep}. Accordingly, the theory is non-Hermitian but
$\mathcal{PT}$-symmetric and thus is physically acceptable. However, one of
the essential problems opposing this theory is that the metric operator has
not been obtained so far. In fact, the form in Eq.(\ref{quasi2}) enables us to
apply the conventional tools to calculate its metric operator at least in a
perturbative way. To show this consider consider the non-Hermitian Hamiltonian
in Eq.(\ref{quasi2}). Mostafazadeh generalized the requirement of real spectra
for a non-Hermitian theory to the existence of a positive definite metric
operator $\eta$ such that $\eta_{+}H\eta_{+}^{-1}=H^{\dagger}$
\cite{spect,spect1} with an equivalent hermitian theory $h$ such that

$\rho H\rho^{-1}=$ $h$ with $\rho=\sqrt{\eta}=\exp\left(  -\frac{Q}{2}\right)
$ where $\eta=\exp\left(  -Q\right)  $. Accordingly, we can get%

\begin{align*}
H^{\dagger}  &  =\exp(-Q)H\exp(Q)=H+[-Q,H]+[-Q,[-Q,H]]\\
&  +[-Q,[-Q,[-Q,H]]]+....
\end{align*}
Also, one has a similar expansion for the Hermitian Hamiltonian $h=\exp
(\frac{-Q}{2})H\exp(\frac{Q}{2})$, which will result in a perturpative
expansion for $h$ as
\[
h=h_{0}+gh_{1}+g^{2}h_{2}+.....
\]

Now, we have for $H^{\dagger}$ the expansion;
\begin{align*}
\exp(-Q)H\exp(Q)  &  =H_{0}+gH_{I}+[-Q,H_{0}]+[-Q,gH_{I}]+[-Q,[-Q,H_{0}]]+\\
&  [-Q,[-Q,gH_{I}]]+[-Q,[-Q,[-Q,H_{0}]]+[-Q,[-Q,[-Q,gH_{I}]]...\\
&  =H_{0}+gH_{I}^{\dagger},
\end{align*}
with
\[
Q=Q_{0}+gQ_{1}+g^{2}Q_{2}++g^{3}Q_{3}+......
\]

Thus, we get a set of coupled equations for the operators $Q_{n}$, where the
first few equations are given by
\begin{align}
0 &  =[-Q_{0},H_{0}]\text{ }\Rightarrow\text{ }Q_{0}=0\ \text{ is a good
choice,}\nonumber\\
H_{I}^{\dagger} &  -H_{I}=-2ig\int d^{3}x\psi^{3}=-\frac{1}{2}[-Q_{1}%
,H_{0}],\nonumber\\
0 &  =\frac{1}{2}[-Q_{2},H_{0}]+\frac{1}{2}[-Q_{1},H_{I}]+\frac{1}{3!}%
[Q_{1},[Q_{1},H_{0}]],\nonumber\\
0 &  =\frac{1}{2}[-Q_{3},H_{0}]+\frac{1}{2}[-Q_{2},H_{I}]+\frac{1}{3!}%
[Q_{2},[Q_{1},H_{0}]]\nonumber\\
&  +\frac{1}{3!}[Q_{1},[Q_{2},H_{0}]]+\frac{1}{4!}[-Q_{1},[-Q_{1}%
,[-Q_{1},H_{0}]]]\label{qpert1}\\
&  +\frac{1}{3!}[-Q_{1},[-Q_{1},H_{I}]],\nonumber\\
0 &  =\frac{1}{2}[-Q_{4},H_{0}]+\frac{1}{4}[-Q_{3},H_{I}]+\frac{1}{3!}%
[-Q_{2},[-Q_{2},H_{0}]]\nonumber\\
&  +\frac{1}{5!}[Q_{1},[Q_{1},[Q_{1},[Q_{1},H_{0}]]]]+\frac{1}{3!}%
[-Q_{2},[-Q_{1},H_{I}]\nonumber\\
&  +\frac{1}{3!}[-Q_{1},[-Q_{2},H_{I}]+\frac{1}{4!}[-Q_{1},[-Q_{1}%
,[-Q_{1},H_{I}]]]]\nonumber\\
&  +\frac{1}{8\times4!}[-Q_{1},[-Q_{1},[-Q_{2},H_{0}]]]]\nonumber\\
&  +[-Q_{1},[-Q_{2},[-Q_{1},H_{0}]]]]+[-Q_{2},[-Q_{1},[-Q_{1},H_{0}%
]]]].\nonumber
\end{align}

In fact, this regime has been used before to calculate the $Q$ operator for a
Hamiltonian form that is similar to the effective form of the $-\frac{\lambda
}{4!}\phi^{4}$ theory in Eq.(\ref{quasi}) in $0+1$ dimensions \cite{bendx4q}.
However, one can expect that the equivalent Hermitian theory is
non-renormalizable. Moreover, following the work in Ref.\cite{cop}, one can
obtain the $Q$ operator up to first order for the more simpler $i\phi^{3}$
theory which will take the form;%

\[
Q_{1}=\int\int\int d^{3}xd^{3}yd^{3}z\left(  M_{(xyz)}\Pi(x)\Pi(y)\Pi
(z)+N_{x(yz)}\psi(y)\Pi(x)\psi(z)\right)  ,
\]
where the functions $M_{(xyz)}$ and $N_{x(yz)}$ are defined there. However,
the resulting Hermitian Hamiltonian $h$ have terms for which the coupling has
a negative mass dimension and the situation will be worst in higher orders as
more powers of both the fields $\Pi$ and $\psi$ are appearing in the operator
$Q$ and thus in the Hermitian Hamiltonian $h$. In other words, the
transformation $\rho H\rho^{-1}=$ $h$ does not respect the superficial degree
of divergence \cite{abonon}. In fact, this was the reason that leads the
authors of Ref.\cite{ptsym} to have a Hermitian theory which is incalculable
equivalent to the calculable non-Hermitian $ix^{3}$ theory. Relying on these
analysis, we expect that this is will be the case for the $-\phi^{4}$ theory
and the calculation of the metric operator which depends on the field and its
conjugate momentum is thus meaningless.

Now, we have a theory (the $-\lambda\phi^{4}$ theory) which shows up
interesting behaviors like asymptotic freedom and it seems that it is free
from the hierarchy problem. However, the theory seems to be incalculable as
well. To escape from this puzzle one can seek another metric operator for
which the Hamiltonian is pseudo-Hermitian and leading to an equivalent
Hermitian form though with a wrong sign propagator. In fact, one may gausses
the parity operator. To do that, we take into account that the quasi-particle
field $\psi$ is pseudo scalar and thus
\[
PHP^{-1}=H^{\dag},
\]
where $P$ is the parity operator. Then, one can introduce the operator $\rho$
such that $P=\rho^{2}$. In this case, we have the following transformation
properties%
\[
\rho\psi\rho^{-1}=-i\psi\text{, \ }\rho\Pi\rho^{-1}=i\Pi.\text{\ \ }%
\]
Note that both $P$ and $\rho$ preserve the commutation relations of the field
operators $\left[  \psi(x),\Pi(y)\right]  =i\delta^{3}(x-y)$. In view of this,
one can obtain
\begin{equation}
h=\rho H\rho^{-1}=H=\int d^{3}x\left(  \frac{-1}{2}\left(  \left(  \nabla
\psi\right)  ^{2}+\Pi^{2}+M^{2}\psi^{2}\right)  -i\frac{\lambda\left\vert
B\right\vert }{6}\psi^{3}-\frac{\lambda}{4!}\psi^{4}\right)  .\label{hermn}%
\end{equation}
Note that $h$ is Hermitian but the propagator has an opposite sign to the
conventional $\phi^{4}$ theory. Moreover, all the Feynman diagrams calculated
either with $h$ or $H$ have the same numerical value as it should be. Also,
both $h$ and $H$ are normalizable theories rather than the expected Hermitian
Hamiltonian obtained from $\exp(-Q)H\exp(Q)$, with $Q$ is a functional in
$\psi$ and $\Pi$ fields and calculated from the coupled set in
Eq.(\ref{qpert1}). Now, the Hamiltonian has ghost states. However, one can
attribute this to a new physical meaning. To show this, consider the
propagator $\frac{-i}{p^{2}-M^{2}}$, in multiplying by $M$ we obtain a new
propagator of the form
\[
\frac{-iM}{p^{2}-M^{2}}=\frac{-i}{\frac{p^{2}}{M}-M},
\]
which can be considered as the matter probability amplitude with the new
hypothesis that matter density can be negative or positive. \ Accordingly, the
wrong sign appears in the propagator can be attributed to a particle of
negative mass \textit{i.e.} made of exotic matter, provided that the jump from
the non-relativistic quantum mechanics to the relativistic quantum mechanics
has done via $\frac{p^{2}}{2M}\rightarrow\frac{p^{\mu}p_{\mu}}{2M}$. Or
equivalently, have the klein-Gordon equation of the form;%

\[
\left(  \frac{\nabla^{2}-\frac{\partial^{2}}{\partial t^{2}}}{2m}-\frac{m}%
{2}\right)  \psi=0,
\]
which is the same for both positive and negative values of $m$. In the
presence of interactions, there exists two different Klein-Gordon equations
one for positive $m$ and another one for negative $m$. However, for a negative
mass particle, the quantum field Hamiltonian for the theory under
consideration have the form;%
\begin{equation}
h=\rho H\rho^{-1}=H=\int d^{3}x\left(  \frac{-1}{2}\left(  \frac{\left(
\nabla\psi\right)  ^{2}+\Pi^{2}}{m}+m\psi^{2}\right)  -i\frac{\lambda
\left\vert B\right\vert }{6}\psi^{3}-\frac{\lambda}{4!}\psi^{4}\right)  ,
\end{equation}
and in this way the negative sign can attributed to a theory of negative mass.
Note that, this form is Hermitian and thus the theory is unitary in the Dirac
sense inner product.

In conclusion, we showed that the non-Hermitian and $\mathcal{PT}$-symmetric
$\phi^{4}$ theory has very interesting features as an asymptotically free
theory. The most important feature of the theory is that the parameters of the
theory are perturbative for UV scales rather than the corresponding Hermitian
theory used to break the $SU(2)\times U(1)$ symmetry in the standard model.
This suggests that using the negative coupling $\phi^{4}$ theory instead, will
solve many problems in the standard model. Out of this problems, is that the
negative coupling $\phi^{4}$ is not trivial as it has interactions allover the
energy scale because of the asymptotic freedom property. Also, it might save
the standard model if the experiments in the LHC was not able to detect the
Higgs particle in a direct manner. Our reasoning is that in this picture the
Higgs particle is a strongly interacting particle and need an infinite amount
of energy to be isolated. Thus, according to this picture, the search of the
Higgs has to be twisted to go the same way we feel the gluons.

A note to be mentioned is that this work does not give a final answer about
the solution of the standard model problems like the Hierarchy and triviality
problems. This is because, the Higgs mass receives other corrections from the
vector Boson fields coupled to the Higgs field and the top quark contribution
should also be taken into account. Taking this into account yield the result
\cite{DJOUADI}
\begin{equation}
M_{H}^{2}=M_{0}^{2}-\frac{\Lambda^{2}}{8\pi^{2}v^{2}}[M_{H}^{2}+2M_{w}%
^{2}+M_{Z}^{2}-4M_{t}^{2}],
\end{equation}
where $M_{0}$ is the bare mass contained in the unrenormalized Lagrangian. By
the renormalization group analysis mentioned above, we made sure that the
first term will be small as $\Lambda$ goes to higher values. \ For other
terms, we know that all the masses in the standard model depend on the vacuum
condensate which has been shown to have an exponential sharp decrease near
$\lambda\rightarrow0^{+}$ \cite{onep}. Accordingly, one may claim that the
Higgs mass will stay protected against perturbations even for high energy
scales. However, this claim should be tested in view of the renormalization
group functions for the other sectors. In fact, this will take a substantial
amount of time but it naturally becomes a target of our future work. The main
aim we wanted by this work to spread the message that non-Hermitian and
$\mathcal{PT}$-Symmetric $\phi^{4}$ as now a physically acceptable model may
help in the solution of the genuine existing problems in the standard model.
These problems are well known to exist because the Hermitian Higgs mechanism
used is both trivial and suffers from the gauge Hierarchy problem. Another
important message that we need to mention is the that the vacuum energy of the
non-Hermitian and $\mathcal{PT}$-Symmetric $\phi^{4}$ is tiny and negative in
$1+1$ dimensions compared to the Hermitian one \cite{abophi41p1,thesis}. In
another work \cite{cern}, we showed that in $2+1$ dimensions, the vacuum
energy is tiny as well as positive for a wide range of energy scales. In fact,
these features are very interesting regarding the very false result of the
prediction of the cosmological constant from quantum field theory which again
is a manifestation of the gauge Hierarchy problem. In fact, positiveness of
the vacuum energy is essential as it describes an expanding universe (
negative pressure). We hope that using the asymptotically free non-Hermitian
and $\mathcal{PT}$-symmetric $\phi^{4}$ theory in the standard model will
solve such genuine problems relying on the interesting features we explored
above. 
\newpage

Since up till now the metric operator is not known, we tried to give the
negative norm a physical meaning by considering the Higgs mass as a charge
which can be positive or negative. We assert that, the form $\eta=\exp(-Q)$
can be obtained perturbatively using effective field representation. However,
we did not prefer this direction as the resulting Hermitian Hamiltonian will
be non-Renormalizable and thus incalculable.

A Higgs particle with negative mass makes sense in understanding how a
potential bounded from above can have stable states. In this case, negative
mass particles have the property of maximizing the classical action rather
than the conventional positive mass particles which minimizing the classical
action. In other words, a negative mass particle feels the bounded from above
potential as the positive mass particle feels the bounded from below potentials.

\newpage


\end{document}